# LB2CO: A SEMANTIC ONTOLOGY FRAMEWORK FOR B2C ECOMMERCE TRANSACTION ON THE INTERNET

Akanbi Adeyinka K.

*Institute of Science & Technology, Jawaharlal Nehru Technological University, Hyderabad 500085, A.P, India.*
*Email: akanbiadeyinka@hotmail.com*

*Abstract: Business ontology can enhance the successful development of complex enterprise system; this is being achieved through knowledge sharing and the ease of communication between every entity in the domain. Through human semantic interaction with the web resources, machines to interpret the data published in a machine interpretable form under web. However, the theoretical practice of business ontology in eCommerce domain is quite a few especially in the section of electronic transaction, and the various techniques used to obtain efficient communication across spheres are error prone and are not always guaranteed to be efficient in obtaining desired result due to poor semantic integration between entities. To overcome the poor semantic integration this research focuses on proposed ontology called LB2CO, which combines the framework of IDEF5 & SNAP as an analysis tool, for automated recommendation of product and services and create effective ontological framework for B2C transaction & communication across different business domains that facilitates the interoperability & integration of B2C transactions over the web.*

## I. INTRODUCTION

The popularity of the internet and the huge growth of new internet technologies have led in the last decade to the creation of great amount of eCommerce applications7. The context of B2C eCommerce application requires that an effective communication between the machines is possible. In other words, semantic interoperability between the information systems involved in the communication is crucial [1].

As online shopping has become an important part in people's lives in this 21st century, the product information retrieval mechanism is becoming more and more important, as well as secure communication amongst domains. Information retrieval is the most frequently used method to obtain information in the web, and the purchaser must get access to the product information before a transaction could be performed.

However, the current way of information organization and expression is defective, in that it was designed for user decision making needs, rather than to provide semantic information that computer can process automatically, thereby limiting the computer's capacity of automatic analysis and further intelligent process in information retrieval, that enhances information and data sharing across various platforms.

However, two extremely important factors that can contribute to this effective non-human communication are: (1) a common language in which the resources implied in the communication can be specified, and (2) a shared knowledge framework and vocabulary between the different systems that are present in the eCommerce domain8. They are syntactic and semantic dimensions. The first, syntactic dimension has led to the creation of varied representation languages for the specification of web resources (XOL, SHOE, RDF, RDF Schema, OIL and DAML+OIL). The semantic dimension is related with the knowledge framework and vocabulary used by the systems involved in the communication. Therefore, the use of a shared and common knowledge framework and vocabulary increases the interoperability among existing and future eCommerce systems [2].

The major motive of this research is to develop and propose an ontology framework that will facilitate the interoperability and integration of eCommerce transaction on the Internet, by focusing on the semi-automatic integration of existing standards and initiatives in a multilayered eCommerce knowledge model for eCommerce applications through ontology evolution. Ontology evolution is the process that leads to the creation of new ontological models to accommodate future modification of ontology. This process may lead to changes on the design of the ontology, which must be implemented carefully and tailored carefully to achieve the desired outcome. In this research, we employ two promising semantic ontologies SNAP and IDEF5, for comparison and analysis, and the limitations mitigating against this existing model are removed. Thus, using the process of ontology evolution to create the standard ontology framework, which is the aim of this research.

The motivation for this evaluation is two-fold. On the one hand there is need to understand the similarities and differences between the two ontologies and thus enhance the understanding of what eCommerce ontological framework actually are, On the other hand the second aim is to integrate this two ontologies in order to improve the representation, design, analysis, & interoperability of eCommerce ontology framework in the domain. The expected result is a multi-layered ontological framework called LB2CO, which present a graphical and structured ontology, which is independent across different business domains, gives rich semantic relationships among entities in a domain for easy searching and communication, and can be implemented using different languages to allow the interoperability of vertical markets and integration between different B2C eCommerce transactions.

The first implementation of LB2CO focuses on online eCommerce business model, by specifically allowing developers to create "ontologydriven" B2C eCommerce websites. The framework is demonstrated with the "Semantic AUTO Store", which uses the LB2CO ontology for effective searching & interoperability across different domains. In section 2, related terminologies and limitations in the current eCommerce model are discussed. Section 3 discusses SNAP & IDEF5, Section 4 introduces the architecture of proposed LB2CO semantic ontology for eCommerce applications, and the case study is explained in section 5. Finally the conclusion is given.

## II.  RELATED TERMINOLOGY

### A.  E-Commerce

Electronic Commerce or eCommerce can be defined as the exchange of goods and services by means of the Internet or other computer network infrastructures. eCommerce follows the same basic principles as traditional commerce—that is, buyers and sellers come together to exchange goods for money. In eCommerce, buyers and sellers transact business over networked computers, which can be across cities, countries or continents. There are two major eCommerce styles, they are: Business-2-Consumer and Business-2-Business eCommerce models.

The B2C models operation is the one that uses the Internet to sell products or services directly to consumers or end users. In the B2C eCommerce the Internet and particularly the web is the medium for marketing, sale and post POS channel. The B2B eCommerce model involves Companies doing business with each other such as: manufacturers selling to distributors and wholesalers selling to retailers. In this research effort is concentrated only on B2C ecommerce.

### B.  Current E-Commerce

A search for any product offers is the starting point for most eCommerce transactions. ECommerce web applications are designed to return the most appropriate data to the user based on limited keywords supplied by the user, and the current applications are failing in returning the relevant data to the consumers.

#### Limitations in the Current E-Commerce are:

- Interoperability in an inconsistent environment: This situation occurs where the consumer is in the conflicting state to choose the best option from the available websites.
- Information retrieval & search Disparity: This situation occurs when the machine cannot intelligently recommend products based on existing search indexes.

### C.  Semantic Web

The Semantic Web is not a separate web but an extension of the current one, in which the semantics of information and the services of the web is defined, making it possible for the web to understand and satisfy the requirements of the people to use the web content [3, 4], better enabling computers and people to work in cooperation. To make the web semantic, there is a need for new standard web ontology languages. Ontology is a key and prerequisite for a working semantic web. Ontology's are used to express information in a machine interpretable form, but due to the early developmental stages of the semantic web, many people are not interested in producing ontology. One way of overcoming this problem is to semi-automatically create business ontology from existing resources like knowledge base model, to enhance the rapid development of semantic web.

### D.  Ontology

Ontologies can be defined as "formal and explicit specifications of a shared conceptualization". Ontologies are central to the implementation of the Semantic Web. They contain domain knowledge, specific data regarding a certain subject field, in a very structured way, if we compare this definition with the one given for the Semantic Web in [5], "the conceptual structuring of the web in an explicit machine-readable way". It helps to achieve interoperability and communication among software systems, improve the design and quality of software systems and play a key role in agent communication. As it improves the accuracy of searching and enables the development of powerful applications that tackle complicated queries, whose answers do not reside on a single web page. Some basic ontology languages are XML/XML Schemas, RDF and RDF Schemas.

## E. XML/XML Schemas

XML (Extensible Markup Language) is a formal language that conforms to the SGML specifications. It can be seen as a subset of SGML, which is simpler and more practical in its use than SGML. XML enables clear unambiguous data representation with well-defined syntactic means. While XML is highly helpful for a syntactic interoperability and integration, it carries as much semantics as HTML. Nevertheless, XML solved many problems, which have earlier been impossible to solve using HTML, that is, data exchange and integration in a well précised manner. A well-formed XML document creates a balanced tree of nested sets of open and closed tags, each of which can include several attribute-value pairs. The following structure shows an example of an XML document identifying a "Contact" resource. The document includes various metadata markup tags, such as <first_name>, <last_name>, and <email>, which provide various details about a contact.

```
<Contact contact_id="033220">
<first_name> Arun</first_name>
<last_name> Kumar </last_name>
<college> Jawaharlal Nehru Technological University
</college>
<state> Andhra Pradesh </state>
<country> India </country>
<email> akanbiadeyinka@hotmail.com </email>
<phone> 91********* </phone>
</Contact>
```

*Code Snippet 1: XML Schemas*

## F. RDF (Resource Description Framework)

RDF is a language for expressing data models in XML syntax. XML provides an elemental syntax to structure the data [6]. It provides the meaning to that structured data [4,5]. RDF is used to describe web resources. RDF uses XML and it is at the base of semantic web, so that all other languages corresponding to the upper layers are built on it. RDF is a formal data model for machine understandable metadata used to provide standard descriptions of web resources. RDF assertion consists of a triplet subject, predicate, object in which a subject has a property that property value can be either a string literal or a reference to another resource. With RDF it is possible to add predefined modeling primitives for expressing semantics of data to a document without making any assumptions about the structure of the document. RDF defines a resource as any object that is uniquely identifiable by a Uniform Resource Identifier (URI).

## G. RDF Schema

The RDF Schema (RDFS) provides a type system for RDF. The RDFS is technologically advanced compared to RDF since it provides a way of building an object model from which the actual data is referenced and which tells us what things really mean. Briefly, the RDF schema (RDFS) allows users to define resources with classes, properties, and values. A class is a structure of similar things and inheritance is allowed [8]. This allows resources to be defined as instances of classes, and subclasses of classes.

```
<?xml version="2 .0"?>
<rdf:RDF
xmlns:rdf= "http://www.w.org/ 999/0 / -rdf-syntax-ns#"
xmlns:rdfs="http://www.w.org/ 000/0 /rdf-schema#"
xml:base= "http://www.hr.com/humanresources#">
<rdf:Description rdf:ID="faculty">
<rdf:type
rdf:resource="http://www.w.org/000/0
        /rdfschema#Class"/>
</rdf:Description>
<rdf:Description rdf:ID="Associate Professor">
<rdf:type
rdf:resource="http://www.w.org/000/0
        /rdfschema#Class"/>
<rdfs:subClassOf rdf:resource="#faculty"/>
</rdf:Description>
</rdf:RDF>
```

*Code Snippet 2: Resource Description Framework Schemas (RDFS)*

## H. Components of Ontology

There have been different representation & formalization of ontologies. Each of which incorporates different components that is used during ontological processes and task execution. However, they share the following minimal set of components namely:

- *Classes:* This represents concepts, within a specified domain. For instance, in the tourism domain, concepts are: locations (cities, villages, etc.), lodgings (hotels, camping, etc.) and means of transport (planes, trains, cars, yacht, and ships). Classes in the ontology are usually organized based on the level of semantic used, where they are interrelated through class inheritance.

- *Relations:* Relations represent a type of association between concepts of the domain. They are formally defined as any subset of a product of n sets, that is: R $\subset$ C1 x C2 x ... x Cn. Ontologies usually contain binary relations. The first argument is known as the domain of the relation, and the second argument is the range. For instance, the binary relation arrivalPlace has the concept Travel as its domain and the concept Location as its range.

- Instances: Instances are used to represent elements or individuals in the domain of the ontology [13].

## I. Level of Semantics

- Semantics is the study of the meaning of signs, such as terms or words. Depending on the models, or methods used to add semantics to terms, different degrees of semantics can be achieved. There are four levels of semantic representations that can be used to semantically describe terms, they are, controlled vocabularies, taxonomies, thesaurus, and ontologies [14]. These four model representations are illustrated in below.

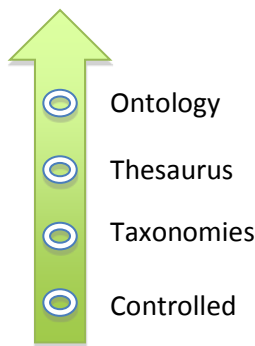

*Figure 1: Levels of Semantic*

- *Controlled Vocabulary:* Controlled vocabularies are at the lowest level of the semantic spectrum. A controlled vocabulary is a list of terms. All terms in a controlled vocabulary should have an explicit, non-redundant definition and are the lowest level of classification. A controlled vocabulary is the simplest of all metadata methods of classification and it's the most commonly used method of classification. For example, flipkart.com has the following of controlled vocabulary below.

*Table 1: Controlled Vocabulary*

| Books | Electronics | Travel |
|---|---|---|
| Comic Books | Camera & Photo | Motorcycle |
| Educational | Books | Television Outlet |
| Novels | Mobile Phones | Auctions |
| Yellow Pages | Jewelry & Watches | Automotive |

- *Taxonomies:* Taxonomy is a category-based classification that arranges the terms in a controlled vocabulary into a hierarchy. Taxonomy classifies the terms in the shape of a hierarchy or tree such as a subset of class. It describes a word by making explicit its relationship with other terms. The hierarchy of taxonomy contains parent-child relationships

- *Thesaurus:* A thesaurus is a networked collection of controlled vocabulary terms with basic relationships between terms. A thesaurus is an extension of taxonomy by allowing terms to be arranged in a hierarchy and also allowing other statements and relationships to be made about the terms. [15] Taxonomies: Taxonomy is a category-based classification that arranges the terms in a controlled vocabulary into a hierarchy. Taxonomy classifies the terms in the shape of a hierarchy or tree such as a subset of class. It describes a word by making explicit its relationship with other words. The hierarchy of taxonomy contains parent-child relationships

- *Thesaurus:* A thesaurus is a networked collection of controlled vocabulary terms with basic relationships between terms. A thesaurus is an extension of taxonomy by allowing terms to be arranged in a hierarchy and also allowing other statements and relationships to be made about the terms. [15]

### III. METHODOLOGY

In the previous section the limitations of the current eCommerce and its drawbacks are stated and how the Semantic web ontology can overcome these drawbacks. The section deals with the development of the proposed LB2CO semantic ontology framework for eCommerce transactions.

*A. Semantic Web Ecommerce Architecture*

The figure below shows the architecture of the semantic web based eCommerce application. The Producer manufactures the products and advertises the details in the web market. Consumer is an individual or end user who buys products or services for personal use over the Internet or over networked connections. Agents are meant to reduce the consumer's work and information overload [7]. In the increasing growth of eCommerce technology, services and information available on the Internet, an agent plays a very important role. Agents are active personalized software's to which tasks can be delegated.

In this semantic architecture, we have two types of agents, namely:

1. Search agent
2. Ontology agent

The consumer directly communicates with search agent, either through the search box; the agent is responsible for retrieving the metadata of documents, based on user defined keyword inputs. The Producer communicates with ontology agent, who provides the knowledge of ontology to answer queries about the domain and its structure.

From the architecture it has shown that, any product or service should be described ontologically to retrieve the result in semantic manner. For example, the company provides the related terms and reference related to the database of the domain of the agent. The agent generates the RDF based on the user search query. Therefore, whenever the consumer or user wants to search for information, the search query is passed to the search agent; the search agent searches the related information based on the ontology model.

## B. Ontology Design Process

There are currently many methods used for building ontologies for eCommerce transactions on the Internet. To develop ontologies in eCommerce, there is a need for a tool to analyze the case of study, while enhancing effectiveness and reducing its limitations. In this research, comparison of SNAP and IDEF5 is done to ultimately arrive at a more comprehensive ontology called LB2CO for the design and analysis of business models for the eCommerce. The aim is at identifying the similarities and differences of both business models in order to merge and integrate them, thus eliminating the limitation. However, this leads to further research to connect both ontologies, such that both SNAP and IDEF5 are employ for the design, schematics representation and integration of the business models into semantic web. However, it is necessary to give an insight into both existing ontologies, to know their differences and utilize their effectiveness. Only then it is possible to produce a consistent and well-related overall ontological framework.

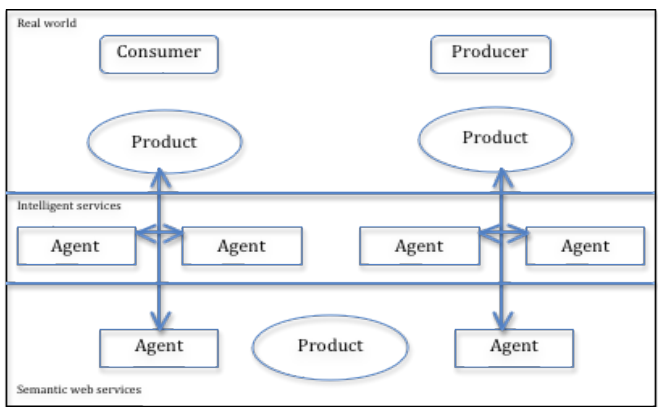

*Figure 2: Semantic Architecture showing role of Ontology Agents.*

Most methods used in ontology design process tends to balance and enhance ontology models through the incorporation of existing standards or simply by building an ontology by eliminating the limitations of existing ontological models [12]. This section explains the main steps of the method used for building eCommerce ontologies, with insight into SNAP and IDEF5 as a major tool for the ontology development process.

- *Selection of standards, joint initiatives, laws, etc., of classification of products and services:* Standards, joint initiatives, laws are a good starting point for the creation of ontologies, since they are pieces of information that have been agreed by consensus or are followed by a community or domain & provide a commonly agreed taxonomy of products and/or services. Several proposals that have arisen, in the context of the eCommerce domain, for the classification of products: UNSPSC, RosettaNet and

e-cl@ss. These initiatives are being developed to ease the information exchange between customers and suppliers.

- *Enrichment of the integrated Ontology:* Current ontology standards do not include detailed attributes of products, relations between products & effectiveness of search queries. They are just categorized using taxonomies and thesaurus. They can be enriched further with information through the use of detailed metadata, such as provided by using XML.

- *Design of multi-layered knowledge architecture:* This step embroils taking into account the main features of the selected sources of information for the particular domain, the aim of this step, is for the identification of relationships between components in the different taxonomies.

- *Knowledge models extraction:* This step involves automating the process of knowledge acquisition from the sources of information previously selected by taxonomy, adapting them to the knowledge model, which can be then represented in XML or RDF schema, and using its import functionality to upload them into the domain platform.

- *Integration of knowledge models:* The knowledge models that have been represented by XML or RDF platform are integrated in the layered architecture, using the semantic relationships identified at the design phase [9].

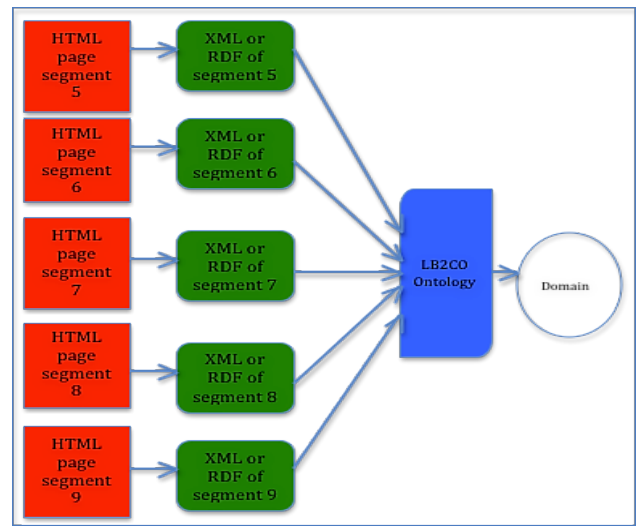

*Figure 3: eCommerce model of LB2CO*

## C. SNAP (Situation, Needs, Actions and Plans)

This is eCommerce model developed for an automated system for recommending products and services to consumers. The automated system was originally developed for the domains of financial planning and banking and has since been extended for insurance, eCommerce telephony applications. SNAP utilizes two sorts of relations: basic relations, and derived relations, which are built out of basic relations using the construction operators of regular expressions. These derived relations allow great expressivity. In contrast to many eCommerce ontologies, which are primarily organized around the concept of product and service, SNAP is based on a commonsense theory of agent interaction between user & the domain.

The basic concepts of SNAP are based on AI theories, and integrated with the concepts of situation, fluent, and actions towards achieving a target goal.

*Situations and fluent:* A situation is a time slice of the world: it describes the way the world is at a particular moment in time. As in the situation/fluent calculus, we speak of a fluent $f$ being true in situation s—Holds ($s,f$) to capture this notion. There are several important types of fluents, enumerated below:

- *Life Stages:* Life Stages depict the fluents that describe some major stage of a domain. e.g A person's life. Age life stages, Career stages, and Family stages.

- *Demographics:* These include such facts as marital status, income, and address.

- *Life Style:* These include a person's habits, such as living expensively, or high-class, middle-class & lower class.

- *Obligations:* Obligations include financial and non-financial commitments.

- *Needs:* A need represents something useful, which an agent does not have. It is quite similar to the standard AI concept of a goal.

- *Events:* An event is defined as a noteworthy happening or occurrence. Any event can be categorized as either an action or a behavior. Actions are those events, which are planned; behaviors are those events that are observed.

- *Relations:* These are the basic relations and derived relations. The derived relations are composed from the concepts and basic relations. The use of basic or derived relations between events makes SNAP a near perfect ontology with limitation of not explicitly representing and reasoning about multiple agents.

### D. IDEF 5

IDEF5 ontology development process consists of the following five activities.

- *Organizing and Scoping:* The organizing and scoping activity establishes the purpose, viewpoint, and context for the ontology development project.

- *Data Collection:* During data collection, raw data needed for ontology development is acquired.

- *Data Analysis:* Data analysis involves analyzing the data to facilitate ontology extraction.

- *Initial Ontology Development:* The initial ontology development activity develops a preliminary ontology from the data gathered.

- *Ontology Refinement and Validation:* The ontology is refined and validated the ontology to complete the development process [10].

### E. IDEF5 ONTOLOGY LANGUAGES

Supporting the ontology development process are IDEF5's ontology languages. There are two such languages that are involved in IDEF5 ontological process: the IDEF5 schematic language & IDEF5 elaboration language. The schematic language is a graphical language, specifically used by domain experts to express the most common forms of ontological information in a graphical detail manner using the construct below in figure 4. This enables average users both to input the basic information needed for a first-cut ontology and to augment or revise existing ontologies with new information. The other language is the IDEF5 elaboration language, a structured textual language that allows detailed characterization of the elements in the ontology [11].

Various pictorial schematics can be constructed in the IDEF5 Schematic Language. The purpose of these schematics, like that of any graphical depiction, is to represent information in a pictorial format. Thus, semantic rules must be provided for interpreting every possible schematic relationship. However, the character of the semantics for the Schematic Language differs from the character of the semantics for other graphical languages. The reason for this is that the chief purpose of the Schematic Language is to serve as an aid for the construction of ontologies; they are not the primary representational medium for storing them. The Schematic Language is, however, useful for constructing first-cut ontologies.

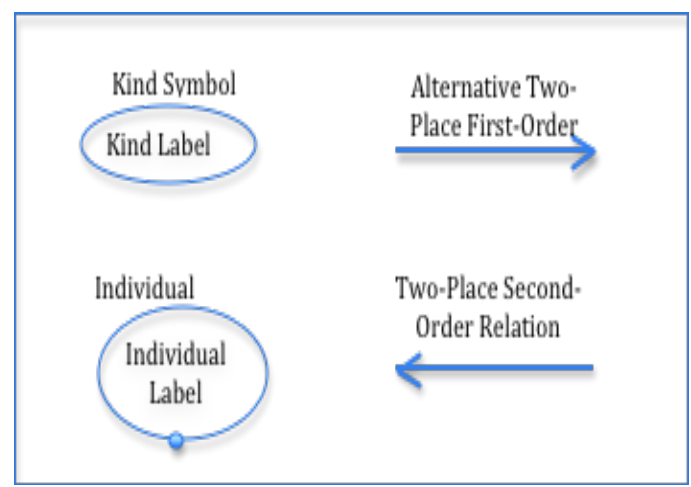

*Figure 4: IDEF5 Schematics*

However, the basic paradigms of the Schematic Language are designed specifically to capture simple but detailed relationship information about real-world objects & their properties.

### F. LB2CO

This is an integrated eCommerce ontology framework that is being developed as the very basis of this research, for an automated system for facilitating B2C eCommerce transactions. It utilizes the basic concept of SNAP and uses IDEF5 ontology development process and thus eliminating their limitation of SNAP & IDEF5 towards the development of enhanced eCommerce Ontological framework.

## IV. CASE STUDY-PROTOTYPE SYSTEM USING LB2CO ONTOLOGY FRAMEWORK

This section aims to define the basic methodology of the proposed eCommerce model literature one step further, by applying a case study approach to show the implementation of the model & framework adopted. The model ontology framework has in some ways been inspired by the different ontology enterprise projects described in academic literature [15]. Ontology essentially gives a common understanding of a specific domain by defining its elements and the relationships between these elements [15].

In this research, an eCommerce experimental prototype website that implements LB2CO ontology development method, the prototype website is a type of eCommerce website that deals with selling of car spare parts & accessories to it customers. This prototype is built to facilitate the illustration of the feasibility and the validity of this framework. This section, demonstrates an application of the prototype. The prototype eCommerce website is designed to apply the concept of business ontology to an actual business case. This case is a typical style of B2C eCommerce.

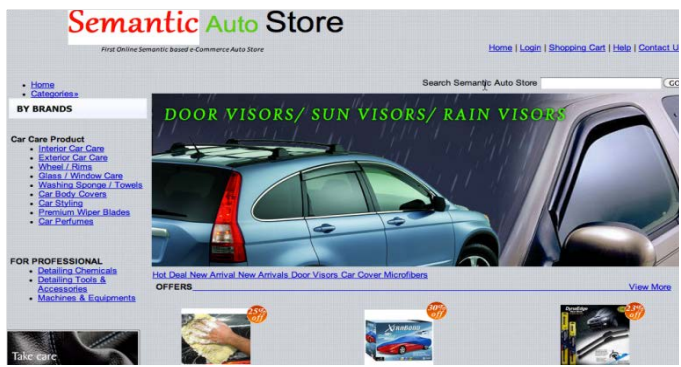

*Figure 5: Semantic Auto Store*

The "Semantic Auto Stores" offered for sale, different spare parts and accessories of automobiles. The knowledge base is used to analyze the interrelations between the different entities. The queries can be either input by the user on the semantic auto stores via the text input box and POST, or they can be passed in via an encoded URL in a GET request or through other forms of input methods or classification on the site.

To develop the ontology LB2CO for the prototype website, the basic essential and elements of eCommerce transaction between the business enterprise and the consumers and also the relationship between this various elements are identified. SNAP methodology will be used to study the rich semantic relationship of the entities involved in the transaction and the inter-relationship between the various entities of the domain.

Afterwards, the Description Summary Form and the Term Description Form are used as tools to show the various entities involved before the application of the LB2CO ontology framework or model. The LB2CO is fully implemented after analyzing the rich semantic relationships between the entities of the domain by finally depicting it in LB2CO semantic language format. The "Semantic Auto Store" demonstrates how ontologies are appropriate as the backend knowledge base to sell the products in the prototype semantic website and how they overcome the limitations of cataloging & recommendation.

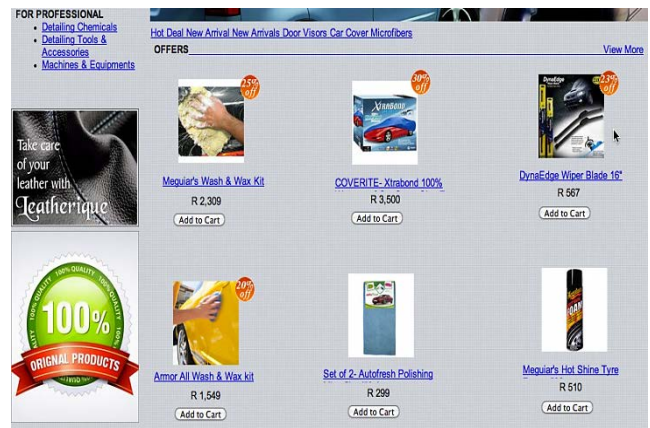

*Figure 6: Semantic Auto Store*

### A. Term Description Entity Form

This form contains the entity terms and their descriptions are fully enumerated below:

*Table 2: Term Description Entity Form for Semantic Auto Store*

| SEMANTIC AUTO STORE ONTOLOGY | |
|---|---|
| PROJECT: - PROTOTYPE eCOMMERCE WEBSITE | ANALYST: - DR. XXXX XXXXXXX |

| | TERM | DESCRIPTION |
|---|---|---|
| 1 | Semantic Auto | The authorized dealer of Semantic Auto Co. |
| 2 | Steering Wheel | An Automobile Steering wheel. |
| 3 | Power Steering Wheel | A variant of steering wheel. |
| 4 | Wiper Blade | An Automobile wiper set |
| 5 | Rims | The automobile rims that is available on sematic auto store. |
| 6 | Door Visor/Sun Visor | The set of automobile visor that is sold on sematic auto store. |
| 7 | Wash & Wax Kit | An Automobile washing kit |

### B. Description Summary Form

The Description Summary form contains the semantic syntax for project description purposes.

*Table 3: Description Summary Form for Semantic Auto Store.*

| Description Summary Form | | |
|---|---|---|
| **Project: Semantic Auto Store Ontology** | Analyst: xxxx xxxxxxx | Reviewer: Prof. xxxx xxxxxx |
| **Version:2.0** | Review Starting Date: | Review completion Date: |
| **Purpose: To develop an ontology framework for Semantic Auto Store** | | |
| **Context: The information acquired must be enough to describe the content of the web page of Semantic Auto Store** | | |
| **Viewpoint: Web Page Visitor** | | |

### V. RESULT & DISCUSSION

The implementation of the LB2CO ontology framework to the latter on Semantic Auto Store provides the inter-operability of this store across different domains on semantic web.

### A. Ontology For The Semantic Auto Store Using LB2CO Schematics

The pictorial representation below shows the Ontology of Semantic Auto Store using LB2CO schematics displaying the rich relationships that's existed between the various entities in the domain. The content of this ontology is completely the same as the ontology that can be presented using other ontology methods. The only difference is the framework adopted for the presentation.

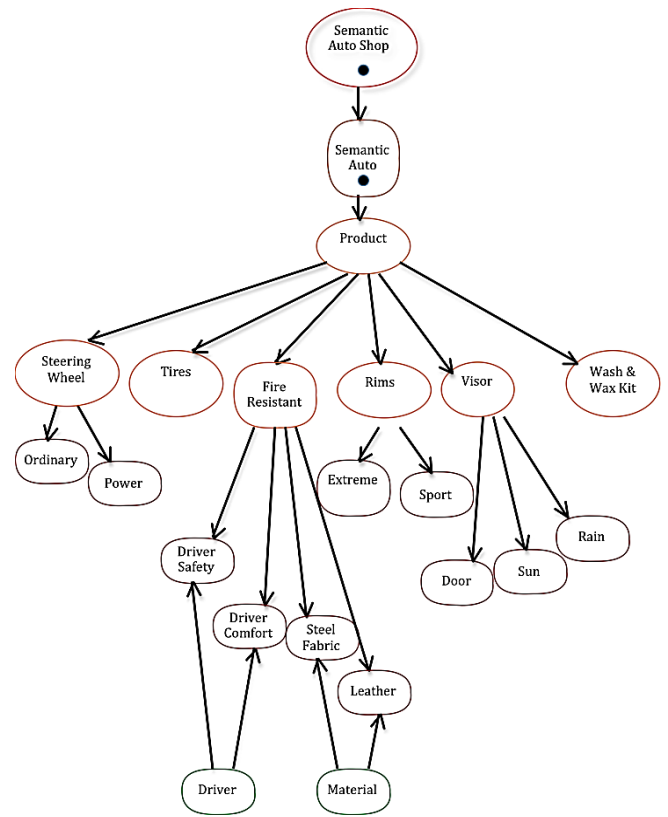

*Figure 7: Using LB2CO Ontological framework Schematics to present the Ontology for Semantic Auto Store.*

### VI. CONCLUSION

In this research, a business ontology framework called LB2CO has been developed for a typical B2C eCommerce transaction. The ontology framework used to develop business ontology for Semantic Auto Store can also be used to develop business ontology for most of small and medium enterprises transactions on the net. However, the fantastic advantage of LB2CO ontological framework is that it can be implemented on different platforms for different domains using different ontological languages.

In this research work, some theoretical methodologies proposed by previous researchers have been verified, and new methodology proposed. Further study in this field can focus on the tangible implementation of this methodology on eCommerce website like Amazon.com etc.